# Electronic and magnetic properties of diiron in extended carbon networks Fe$_2$C$_6$ and Fe$_2$C$_{12}$ from first principles.

Samir F. Matar*

Lebanese German University, Sahel-Alma, Jounieh. Lebanon.

*Formerly at CNRS, University of Bordeaux, ICMCB, 33600 Pessac. France

Email: abouliess@gmail.com & s.matar@lgu.edu.lb

ORCID: https://orcid.org/0000-0001-5419-358X

**Abstract:**

From density functional DFT investigations helped with crystal chemistry rationale, diiron (pairs of Fe), mostly known in molecular diiron nona-carbonyl Fe$_2$(CO)$_9$ and diiron-mono-carbide Fe$_2$C carbide, are embedded in hexagonal C$_6$ substructures. Generated Fe$_2$C$_6$ and Fe$_2$C$_{12}$ are shown to be more cohesive than the mono-carbide on one hand, and increasingly cohesive from hexa- C$_6$ to dodeca- C$_{12}$ on the other hand. From energy differences, the ground state is spin-polarized SP, versus a non-spin-polarized NSP configuration, and identified as ferromagnetic versus a higher energy anti-ferromagnetic hypothesis. The projection of the magnetic charge density on Fe and C, shows that only Fe carries the magnetic moment, while carbon receives charges from Fe as illustrated by the electron localization function ELF 3D and 2D mapping. SP configuration induces an enlarged c/a hexagonal ratio, versus NSP, while a(hex.) remains constant thanks to the rigid C$_6$ carbon substructure network, resulting in an anisotropic magneto-volume response. This feature essentially due to in-plane diiron is discussed from the energy-volume (E, V) NSP and SP equations of state EOS and derived quantities like volume and d(Fe-Fe) change of the magnetization.

# 1-Introduction and context

In paramagnetic elements, a lowering of the energy occurs when the electron spins align, and in certain cases below a critical temperature called the Curie temperature $T_C$, a long-range ferromagnetic order establishes as the ground state. Regarding the highest occupied valence states of the 1st-period transition metals Fe, Co and Ni are ferromagnetic with non-integer magnetic moments: 2.12, 1.69 and 0.86 $\mu_B$ respectively. They are characterized by the onset of magnetic order through "inter d-band spin polarization", implying that the magnetic moments on the 3d sub-shells are mediated by itinerant electrons between them, in a "d….s….d"-like bridging between adjacent sites. The other condition for the onset of magnetic moments is the width of the $n$d band ($n$= 3, 4, 5) which increases along with the $n$ series. The consequence is that mostly 3d metals are magnetically ordered, versus 4d and 5d characterized by a broad d-band and subsequent insufficient localization. Oppositely, the 4f states (and some 5f) are more localized in a narrow band and they keep this behavior in the organized periodic solid. Rare-earth Gadolinium [Xe] $4f^7 5d^1 6s^2$ is characterized by half-filled 4f sub-shell. All 7 electrons polarize to provide a high magnetic moment of 7 $\mu_B$ within a mechanism of "intra f-band spin polarization" [1].

Materials characterized by the presence of pairs of transition metals such as Fe possess particular physical properties due to the interaction of the two metals. Focusing on Fe, diiron compounds are found. The simplest one, ε-Fe is characterized by nearest neighbor distance d(Fe-Fe)=2.47 Å, corresponding closely to the sum of the 2 Fe atomic radii: r(Fe)=1.26 Å. With compounds having more constituents, diiron pairs were early identified in nona-carbonyl $Fe_2(CO)_9$ [3] used in inorganic chemistry synthesis instead of toxic $Fe(CO)_5$. The hexagonal crystal structure contains two diiron pairs with d(Fe-Fe) = 2.52 Å, equal to the sum of two Fe atomic radii, thus highlighting a covalent-like bond. Metal–metal bonding in dinuclear complexes are shown to play a key role in inorganic and organometallic chemistry synthesis [4]. A similar situation occurs in orthorhombic $Fe_2C$ diiron carbide, characterized by d(Fe-Fe) larger than the covalent sum with 2.66 Å [5 (cf. Fig. 1a). Further, theoretical investigations within the well-established quantum density functional theory DFT [6,7] let propose a hexagonal phase of $Fe_2C$ in space group $P6_3/mmc$ (N°194), containing similarly diiron interlayered with carbon [8] (cf. Fig. 1b). However, the carbon network is limited in extension. This observation is relevant given recent investigation within DFT of the stabilization of Fe and Co single atoms in extended carbon hexagonal $C_6$ networks [9]. $FeC_6$ and $CoC_6$ were characterized in the ferromagnetic ground state with integer magnetic moments ($M_{Fe}$ = 2 $\mu_B$, and $M_{Co}$ = 1 $\mu_B$) within an "intra d-band spin polarization", similarly to a 4f element as Gd.

Based on DFT calculations, we consider such $C_6$ based planar networks to examine the magnetic behavior of diiron, in the ground state configurations. Besides the experimentally evidenced $Fe_2C$ diiron carbide, diiron-$C_n$ compounds (n = 6, 12), the effect of the extension of carbon network on the cohesive energy and the magnetic behavior, were considered.

## 2-Computation methodology

At the core of the investigations of new compounds within DFT, the optimization of the candidate structures (the atomic positions and the lattice parameters) onto the ground state is a compulsory step to identify the minimum energy configuration of a given composition. For the purpose, the plane wave Vienna *ab initio* simulation package (VASP) package [10, 11] was used with its implementation of the projector augmented wave (PAW) method [11,12]. The DFT exchange-correlation effects were accounted for with the generalized gradient approximation (GGA) scheme [13]. A conjugate-gradient algorithm according to Press et al. [14] was used to relax the atom positions of the different compositions into the ground-state structure. Structural parameters were considered as fully minimized when forces on the atoms were less than 0.02 eV/Å and the stress components were below 0.003 eV/Å$^3$. Blöchl tetrahedron method [15] was applied for geometry relaxation and total energy calculations. The integrals within the reciprocal space (Brillouin-zone BZ) were approximated using an automatic **k**-point sampling according to the BZ symmetry [16]. The calculations were converged at an energy cut-off of 400 eV for all compounds. The **k**-mesh integration was carried out with increasing BZ precision over successive calculations for best convergence and relaxation to zero strains. Calculations were systematically carried out considering both non-spin-polarized (NSP) and spin-polarized (SP) –magnetic configurations. We also considered an electron localization EL mapping from real-space analysis of EL function (ELF) according to Becke and Edgecomb [17].

## 3-Results and discussion.

### 3.1 Preliminary diiron calculations

ε-Fe is a high-pressure non-magnetic form of iron in the hexagonal system: space group (SG) *P*6$_3$/*mmc*, #194. Iron is in a two-fold 2c Wyckoff position at (1/3, 2/3, ¼), with d(Fe-Fe)= 2.48 Å and V= 20.05 Å$^3$. The electronic structure calculation provides the total energy of E$_{Tot}$=-16.44 eV. Here and in the following calculations, the cohesive energy is obtained by subtraction the energy of Fe in a large box (10×10×10 Å): -7.55 eV from the total electronic energy, leading to E$_{coh}$= –1.34 eV, or E$_{coh}$= –0.67 eV/at. For carbon #194Following this experimental structure, we considered diiron in a similar twofold position: Fe(2d) 1/3, 2/3, ½ but considering the *P*6/*mmm* SG #191 of the model Fe$_2$C$_n$ n= 6, 12 considered herein (cf Table 1). The result after full geometry optimization is E$_{Tot.}$=-14.36 eV and V= 28.62 Å$^3$. The cohesive energy is positive with E$_{coh.}$= +0.74 eV, i.e. a non-bonded diiron system. The result can be explained by considering the small d(Fe-Fe)=2.15 Å separation, and a subsequent smaller volume than ε-Fe. This result is relevant given the stabilization of diiron in *P*6/*mmm* in hexa-carbon C$_6$ networks as shown in the next paragraph.

### 3.2 Diiron carbide Fe$_2$C. Calculations and results

The only diiron carbide is Fe$_2$C crystallizing in the orthorhombic system with *Pnnm* SG #58 with Z= 2 formula units (FU) per unit cell [2]. The structure is shown in Fig. 1a and Table 1 provide the crystal experimental and calculated parameters. The calculations were done in both NSP and SP electronic configuration. A better agreement with the experiment is shown especially for the Fe-Fe distance in the SP magnetic calculations where the magnetization is 6.38 μ$_B$ per 4 Fe. The total and subsequent cohesive energy is larger. However the energy difference is small between the two magnetic configurations, letting suggest that the two configurations are close. Throughout this paper we consider the atom averaged total energy to establish comparisons. Considering the hexagonal form of Fe$_2$C proposed by Fang et al. in 2010 [8] (cf. Fig. 1b), calculations led to a less cohesive system versus experimental orthorhombic diiron mono-carbide. This work will show that by extending the hexagonal carbon network a large stabilization of diiron is obtained.

### 3.3 Diiron in extended honeycomb carbon network.

*a) Structure setups.*

Honeycomb carbon networks were identified in lithium graphitic anode materials $LiC_6$ and $LiC_{12}$ [18] with the *P*6/*mmm* space group. $FeC_6$ was calculated based on $LiC_6$ where we showed the characterization of an integer value of the magnetization of 2 $\mu_B$ [9]. The structure is shown in Fig. 1c with Fe at 0, 0, 0 while C is at C(6k) x,0, ½.
Diiron-$C_6$ shown in Fig. 1d) results from shifting carbon by z= ½ (6j) to have 2 Fe at (2d) 1/3, 2/3, ½ (cf. Table 1b and Fig. 1d). Lastly, $Fe_2C_{12}$ is obtained with the same position for the 2F and C at (12n) x, 0, z (cf. Table 1b and Fig. 1e). Note that these setups are original and not related to those of Li-C [18] although they possess the same *P*6/*mmm* space group.

*b) Calculations and results.*

Table 1b) presents the calculated structure parameters in both NSP and SP configurations. The ground state is identified in SP configuration with a total magnetization of 4.88 $\mu_B$ in $Fe_2C_6$ and slightly reduced to 4.53 $\mu_B$ in $Fe_2C_{12}$ which can be related to the smaller d(Fe-C) involving increased Fe-C bonding in the latter and less free electrons to develop spin polarization. The crystal parameter $x_C$~0.33 is the same in both compounds, which is expected in so far that it rules the $C_6$ planar network. In $Fe_2C_{12}$ the additional $z_C$ ruling the relative position of the $C_6$ layer (cf. Fig. 1e) along with the c hexagonal vertical axis, changes from 0.244 in NSP to 0.236 in SP while c/a ration increases. All these results are consistent in describing a lattice along with *a,b* planar direction i.e. within $C_6$, more rigid than between planes i.e., along *c*: $C_6/Fe2/C_6$. Energetically, the effect of going from the hexa-diiron to the dodeca-diiron is the large increase of the cohesive energy from -1.065 eV to -1.543 eV. Interestingly, the cohesive energy is much larger than the one identified for experimental diiron monocarbide $Fe_2C$.

*c) Magnetic charge density and ELF graphical representations.*

The total magnetization (cf. Table 1)is calculated for the whole chemical system. However, it can be assigned to each atomic constituent upon extracting the atom resolved magnetic charge density. Figure 2 shows sketches of the magnetic charge density envelopes (in yellow) for $Fe_2C$, $Fe_2C_6$ and $Fe_2C_{12}$ identified only on Fe sites. Also, the inter Fe-Fe distances are shown for the minimal calculated values in the SP configurations. While d(Fe-Fe) ~2.60 Å in diiron carbide $Fe_2C$, both hexagonal new diiron –C systems have d(Fe-Fe) ~2.50 Å. It is expected that the magnetization magnitude will be Fe-Fe separation as developed in the next sections for the most cohesive $Fe_2C_{12}$.

Figure 3 shows the electron localization function ELF meant to highlight the different behavior of carbon in orthorhombic $Fe_2C$ versus the hexagonal phases. In Fig. 3a, the 3D volumes of electron localization are observed around C only whereas no localization volumes are to be seen around Fe. This is also observed for Fe pairs in Figs. 3b) and 3c) but the shape of the ELF is different from Fig. 3a) where the electron localization is between carbon atoms forming the $C_6$ network, observed completely in $Fe_2C_{12}$, where the whole two $C_6$ substructures are within the unit cell. The electron localization is further stressed through the ELF 2D slice crossing the Fe-Fe plane, in a supercell 2×2×1 projection for a clear representation. The dominant blue color signals zero electron localization, which is expected from the departure of electrons Fe→C; but between diiron pairs, the localization if of light blue to green color, signaling a nearly free-electron like, electron gas distribution. Then it can be concluded that iron pairs interact through the electron gas as presented in the introduction.

### 3.4 Diiron dodeca-carbide $Fe_2C_{12}$

a) NSP and SP energy-volume equations of state.

Since the ground state diiron–carbon system has been confirmed to be $Fe_2C_{12}$, we further detail its properties by establishing the (E,V) equation of state (EOS) in its two NSP and SP configurations. The physics underlying this procedure is that the calculated total energy corresponds to the cohesion within the crystal in as far as the solution of the Kohn-Sham DFT equations gives the energy for infinitely separated electrons and nuclei. But the zero of energy depends on the choice of the potentials, then energy becomes arbitrary through its shifting, not scaling. However, the energy derivatives, as well as the equation of states, remain unaltered. The NSP and SP E(V) curves in Fig. 4a show a quadratic shape. They are fitted with an energy-volume Birch EOS up to the 3$^{rd}$ order [19]:

$E(V) = E_o(V_o) + [9/8]V_oB_o[[(V_o)/V])^{2/3}-1]^2 + [9/16]B_o(B^{'}-4)V_o[[(V_o)/V])^{2/3}-1]^3$

where $E_o$, $V_o$, $B_o$ and $B'$ are the equilibrium energy, the volume, the bulk modulus, and its pressure derivative, respectively. The fir values are given in the inserts for each magnetic configuration. Oppositely to NSP $Fe_2C_{12}$, the SP ground state is found at lower energy and larger volume. The consequence is found in the bulk modulus $B_0$ which is a measure of the resistance to volume (isotropic) compression. The larger the volume, the more compressible is the chemical system: $B_0(SP) = 250$ GPa $< B_0(NSP) = 300$ GPa with a difference of 50 GPa. From Table 1c, the major difference between NSP and SP is in the c/a hexagonal ratio which is much smaller in NSP.

Such observations lead to consider the relevant magneto-volume effects which are accounted for firstly by the plot of magnetization versus volume shown in Fig. 4b). At low volumes below 90 Å$^3$ magnetization vanishes but the diiron dodeca-carbide remains stable magnetically with magnetization magnitudes from 4 to 5 $\mu_B$ around the equilibrium volume of 121 Å$^3$. Also since the major difference between NSP and SP is in the magnitude of the hexagonal c/a ratio, we show in Fig. 4c the plot of magnetization against c/a ratio, knowing that the equilibrium value is around 1.65. One observes that the effect is of small magnitude (the y-axis is extended to show details). Oppositely in Fig. 4d), the change with the Fe-Fe separation is more significant and the scattered points follow the M(V) curve in Fig. 4b). It can be concluded that $Fe_2C_{12}$ magnetization has an anisotropic response to compression, whereby the main component of magnetization is in-plane. Also in so far that the magnetization remains constant over a broad range of volume and d(Fe-Fe) around minima (cf. Figs. 4b and 4d). The feature can be of relevance upon growing layers on substrates.

    b)  Spin projected density of states

The SP calculations are implicit of ferromagnetic order. Further antiferromagnetic calculations for a check of the ground magnetic state led to an energy increase, letting confirm the ferromagnetic ground state with M= 4.5 $\mu_B$. The corresponding spin projected (↑,↓) total density of states DOS plots are shown in Fig. 5. The lower energy part corresponds to carbon DOS which are equal in position and magnitude for both spin channels (no spin polarization), oppositely to the higher energy part corresponding to Fe DOS which are of unequal weight for the two spin channels: spin polarization and development of Fe magnetic moments. They are crossed by the Fermi level ($E_F$). $E_F$ position is designated by a vertical line, it crosses the Fe at different magnitudes of ↑,↓ finite DOS, letting propose a metallic ferromagnetic compound.

## 4 Concluding notes

The purpose of the present study was to propose new original stable diiron –C compounds with a ferromagnetic ground state. For many years now, the experiment is admittedly helped with theoretical investigations [20], especially using methods based on DFT. It can be expected that targeted experimental investigations are likely to arrive at preparing such diiron compounds for surface coating magnetic applications, using modern deposition methods. Such efforts are planned in collaboration with large universities in Lebanon and in France.

## Acknowledgments

Structure drawings in Figs 1, 2, 3 were done using VESTA graphic software [21].

*Dedication: This work was done during the COVID-19 pandemic confinement using a home computer operating Linux Suse13 on 4 processors. It is dedicated to the memory of the Coronavirus victims all over the world and to the medical and nursing staff as well as to the Red Cross.*

**Tables**

Table 1. Diiron-carbon compounds. Distances are in units of Å (1Å = $10^{-10}$ m). Constituents atoms energies in a large 10×10×10 Å box with spin polarization.
E (Fe -) = –7.55 eV; E( C) = –7.11 eV.

a) Orthorhombic $Fe_2C$ (Z= 2) formula units (FU) [2] and calculated results in non-spin-polarized NSP and spin-polarized SP configurations.

| $Fe_2C$ (Z=2) | Experimental *Pnnm* #58 | NSP | SP |
|---|---|---|---|
| Lattice parameters (Å) | a = 4.711<br>b = 4.318<br>c = 2.830 | a = 4.64<br>b = 4.24<br>c = 2.80 | a = 4.69<br>b = 4.27<br>c = 2.81 |
| | Fe (4g) x, ¼, 0<br>C(2a) 0,0,0<br>x= 0.667 | x= 0.645 | x= 0.665 |
| Distances (Å) | d(Fe-Fe) = 2.66<br>d(Fe-C) = 1.90<br>d(C-C) = 2.83 | d(Fe-Fe) = 2.59<br>d(Fe-C) = 1.90<br>d(C-C) = 2.82 | d(Fe-Fe) = 2.61<br>d(Fe-C) = 1.91<br>d(C-C) = 2.83 |
| Total Energy (eV) | | -50.66 | -51.36 |
| Cohesive Energy (eV) | | -6.24 | -6.94 |
| Cohesive Energy (eV)/at. | | -1.04 | -1.16 |
| Magnetization ($\mu_B$) | | | 6.381 / 4 Fe |

b) Calculated diiron in extended hexagonal carbon networks: $Fe_2C_6$ and $Fe_2C_{12}$ in non-spin-polarized NSP and spin-polarized SP configurations.

| Fe$_2$C$_6$ (Z=1) Space group P6/mmm; #191 | | |
|---|---|---|
| | NSP | SP |
| Lattice parameters (Å) | a = 4.34<br>c/a = 0.88 | a = 4.35<br>c/a = 0.90 |
| Atomic positions | Fe(2d) 1/3, 2/3, ½<br>C(6j) x,0,0; x=0.333 | Fe(2d) 1/3, 2/3, ½<br>C(6j) x,0,0; x=0.332 |
| Distances (Å) | d(Fe-Fe) = 2.51<br>d(Fe-C) = 2.40<br>d(C-C) = 1.45 | d(Fe-Fe) = 2.51<br>d(Fe-C) = 2.45<br>d(C-C) = 1.45 |
| Total Energy (eV) | -65.36 | -66.28 |
| Coh. Energy (eV)/cell | -7.60 | -8.52 |
| Coh. Energy (eV)/at. | -0.95 | -1.065 |
| Magnetization ($\mu_B$) | | 4.88 / 2 Fe |

| Fe$_2$C$_{12}$ (Z=1) Space group P6/mmm; #191 | | |
|---|---|---|
| | NSP | SP |
| Lattice parameters (Å) | a = 4.34<br>c/a = 1.59 | a = 4.33<br>c/a = 1.65 |
| Atomic positions | Fe(2d) 1/3, 2/3, ½<br>C(12n) x,0,z<br>x=0.333; z=0.244 | Fe(2d) 1/3, 2/3, ½<br>C(12n) x,0,z<br>x=0.333; z=0.236 |
| Distances (Å) | d(Fe-Fe) = 2.50<br>d(Fe-C) = 2.29<br>d(C-C) = 1.44 | d(Fe-Fe) = 2.50<br>d(Fe-C) = 2.38<br>d(C-C) = 1.44 |
| Total Energy (eV) | -121.06 | -122.02 |
| Coh. Energy (eV)/cell | -20.64 | -21.60 |
| Coh. Energy (eV)/at. | -1.474 | -1.543 |
| Magnetization ($\mu_B$) | | 4.53 / 2 Fe |

**FIGURES**

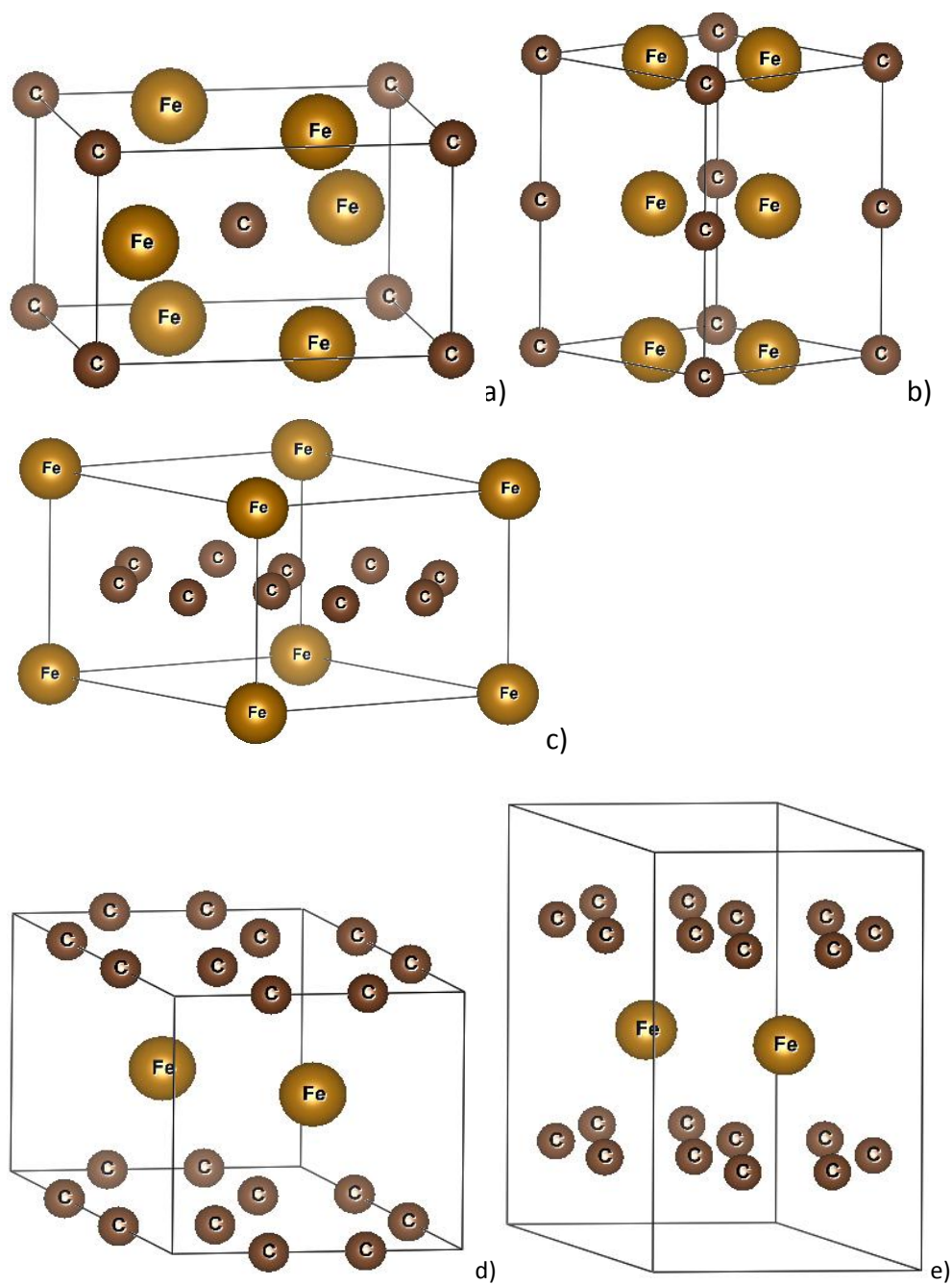

Fig. 1: Structure sketches of di-iron (Fe$_2$) compounds with carbon. a) Orthorhombic Fe$_2$C [2], b) Hexagonal Fe$_2$C [8], c) FeC$_6$ [9], d) Fe$_2$C$_6$ (this work) and e) Fe$_2$C$_{12}$ (this work).

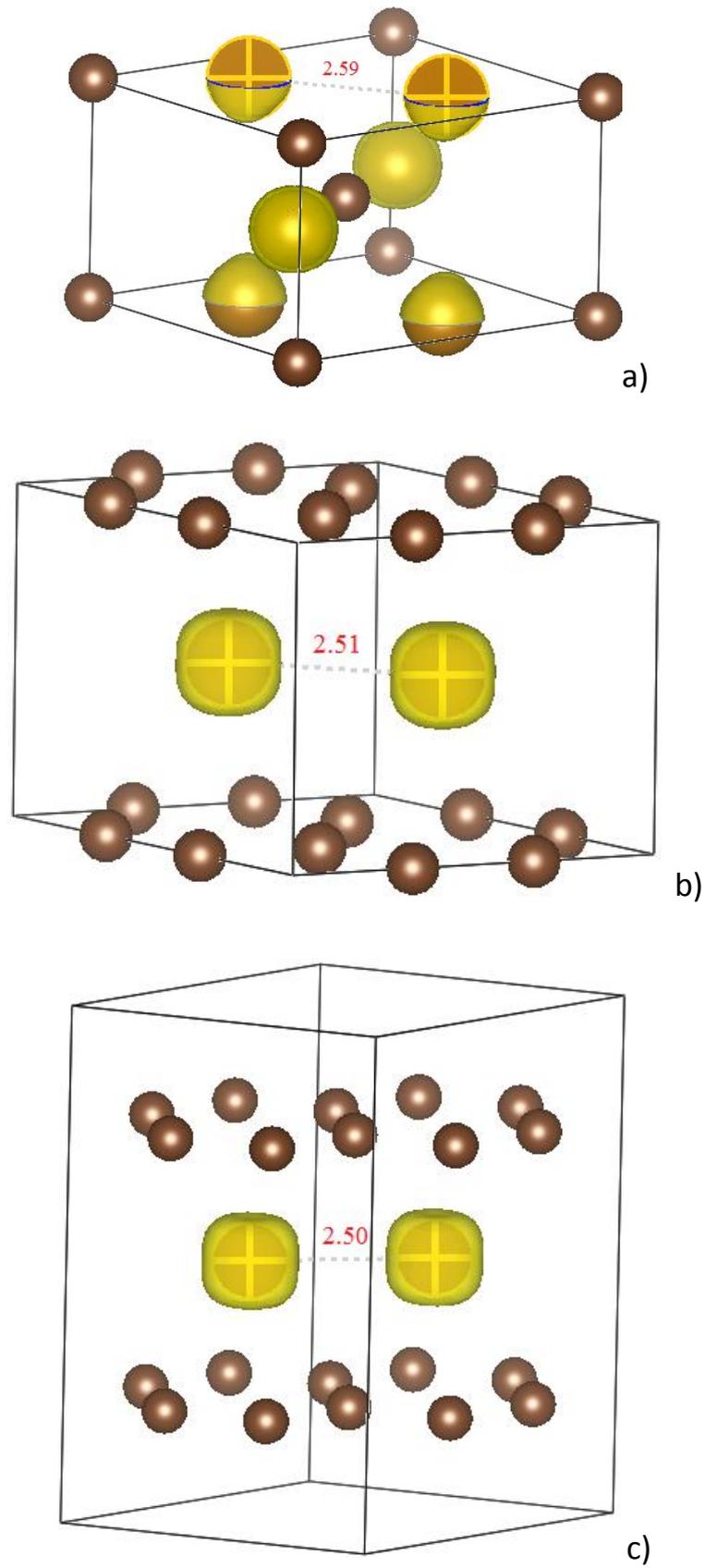

Fig. 2: Magnetic charge density envelopes on Fe in a) Orthorhombic Fe$_2$C, b) Hexagonal Fe$_2$C hyp., c) Fe$_2$C$_6$ and d) Fe$_2$C$_{12}$.

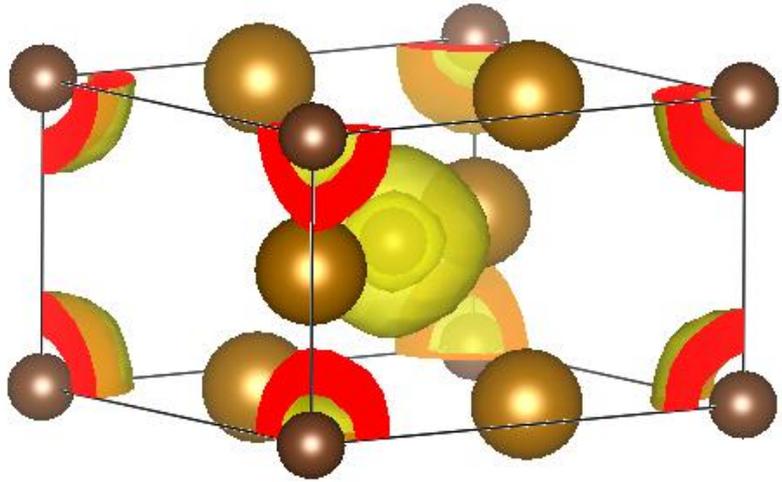

a)Fe$_2$C

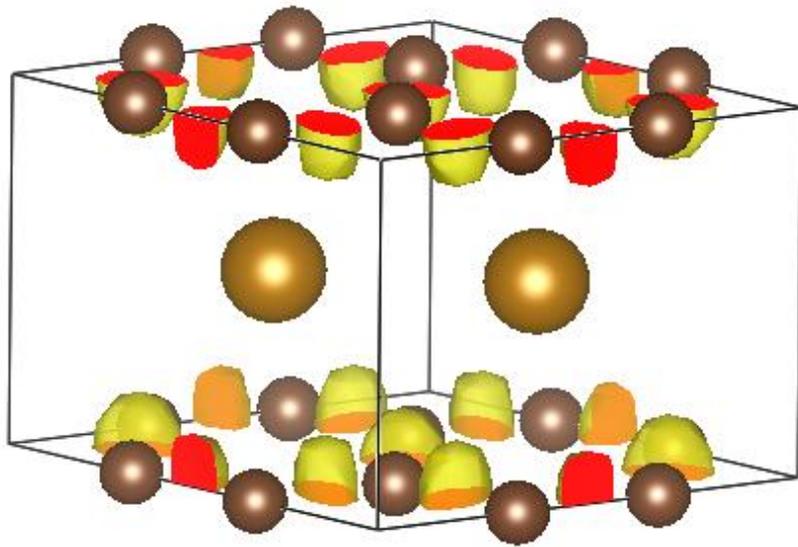

b)Fe$_2$C$_6$

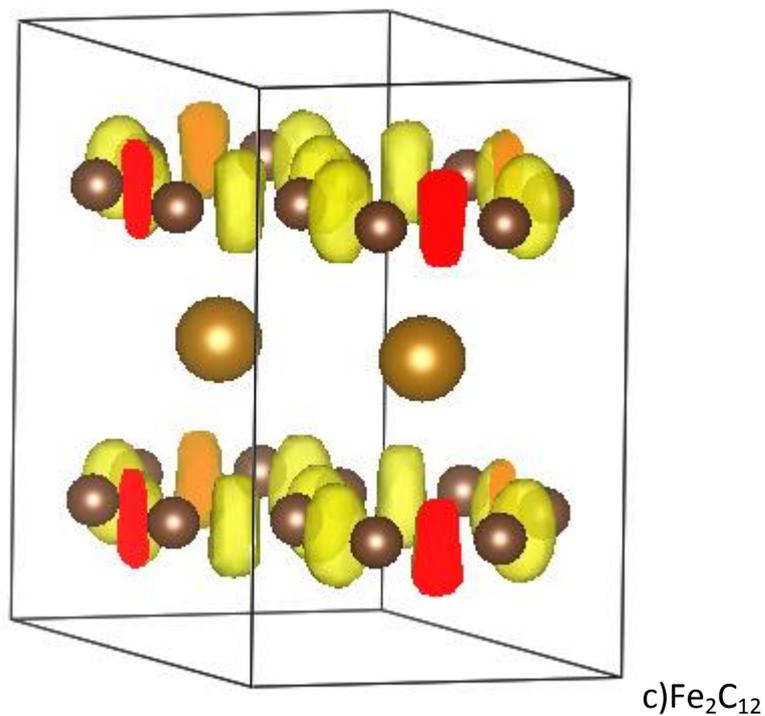

c)Fe$_2$C$_{12}$

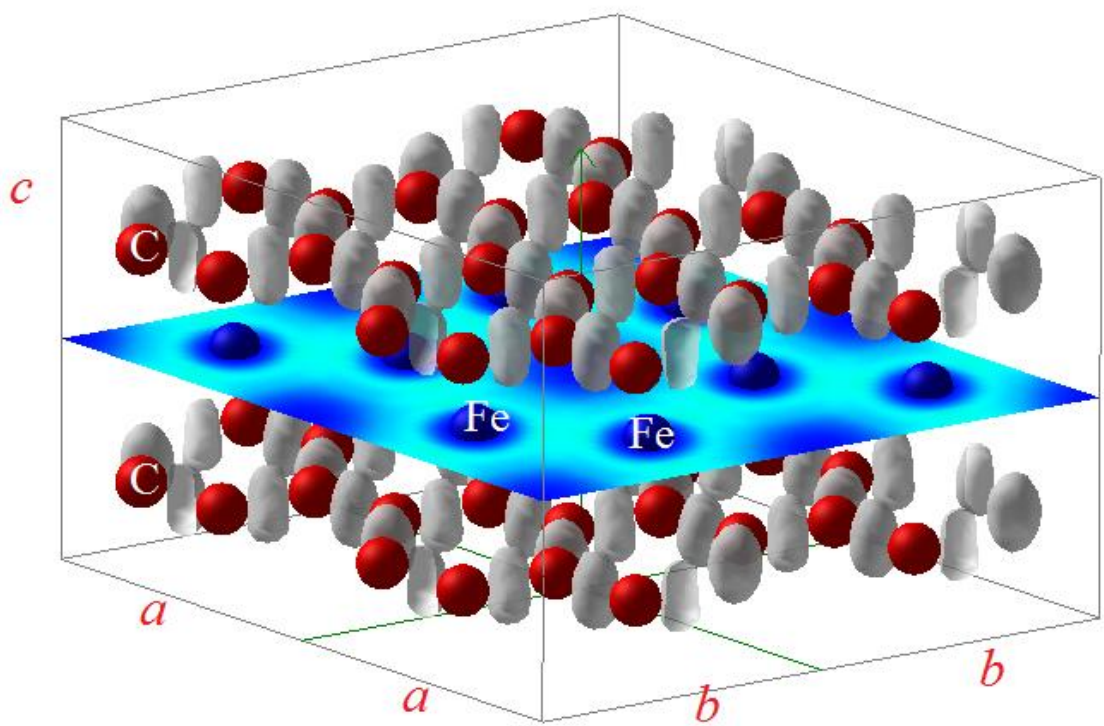

Fig. 3: Electron localization envelopes showing major localization around –a) and b)– and between carbon –c) and d) – atoms in the different diiron-carbon compounds under consideration. Large and small brown spheres correspond to Fe and C respectively. The last panel shows ELF projection over 4 adjacent cells highlighting the diiron planes characterized by nonvanishing localization between two Fe.

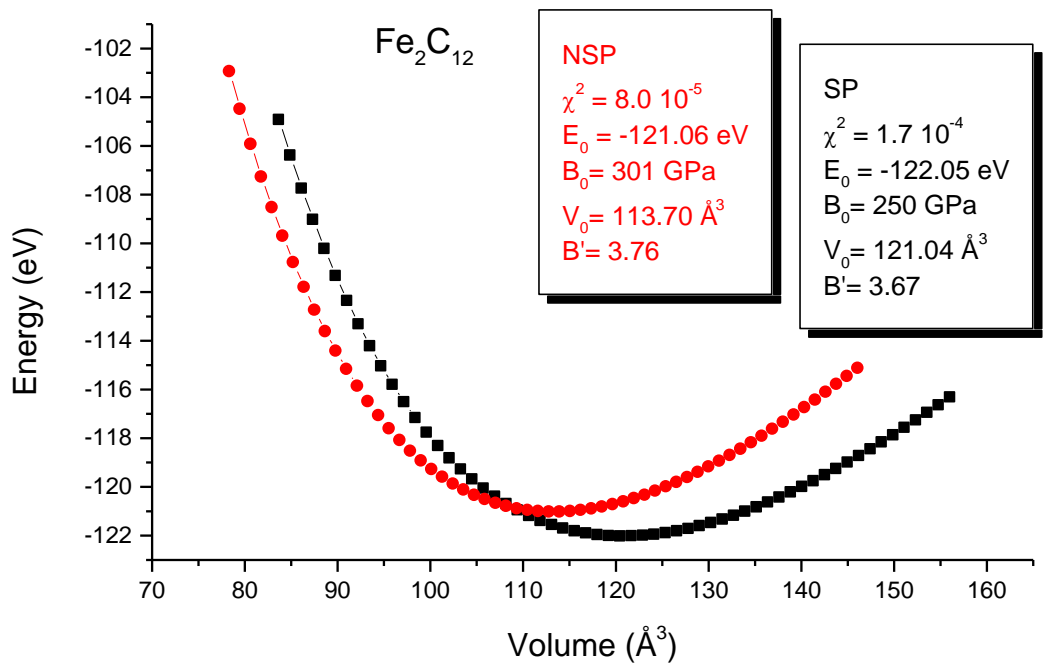

a)

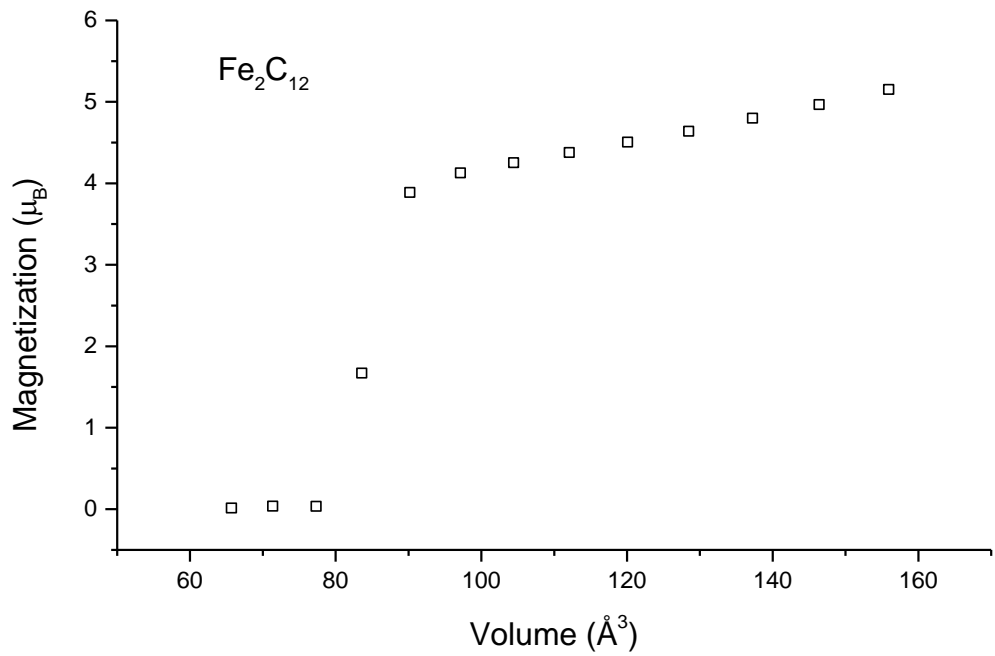

b)

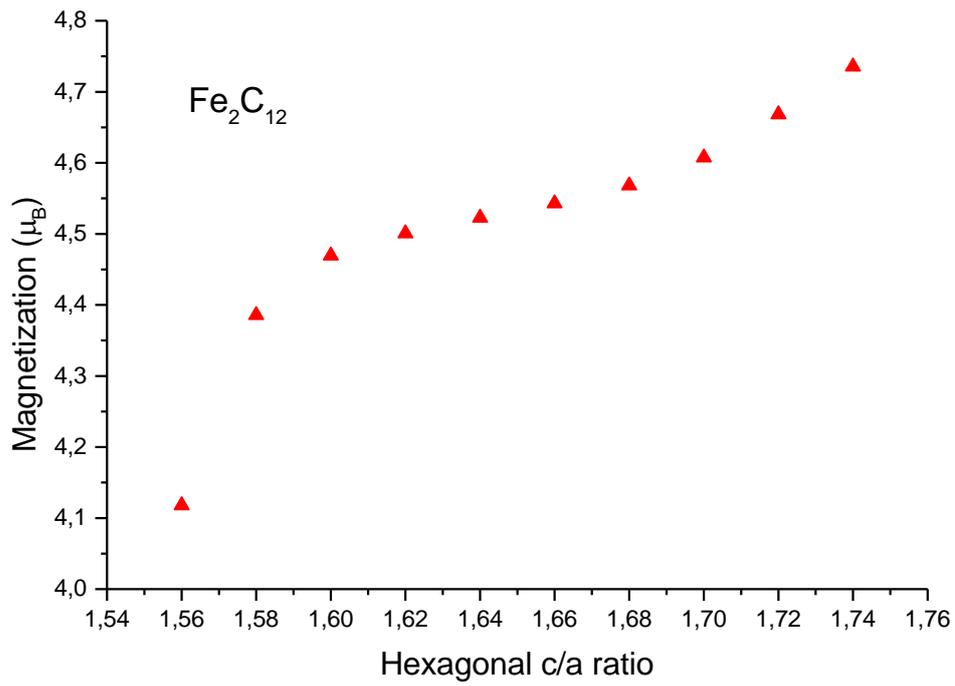

c)

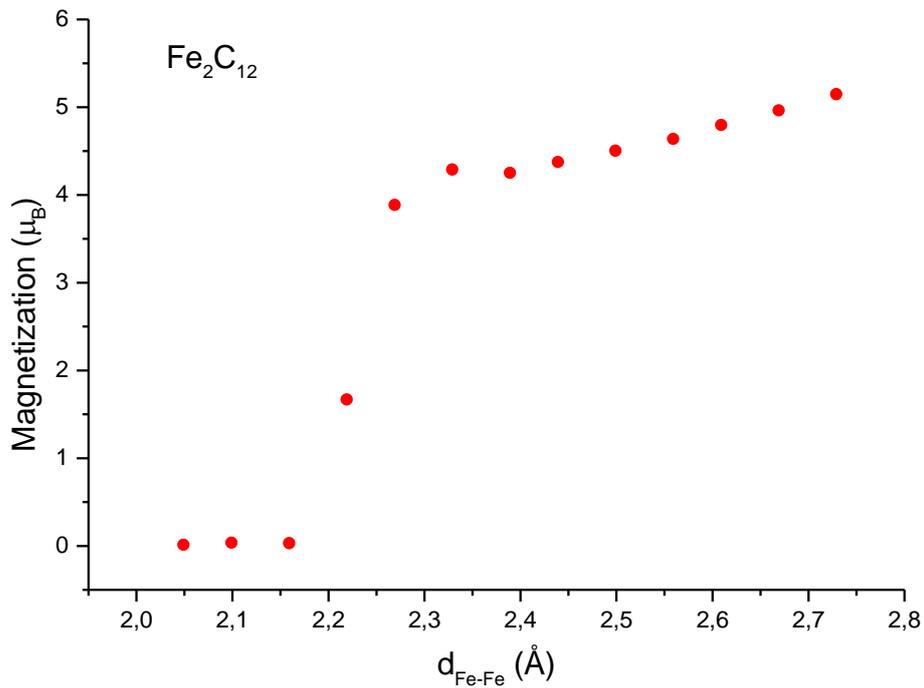

d)

Fig. 4: $Fe_2C_{12}$. a) Energy-volume curves in NSP and SP configurations and Birch EOS fit values in the inserts; b) Magnetization versus volume; c) d(Fe-Fe) change of the magnetization; d) Magnetization versus c/a ratio.

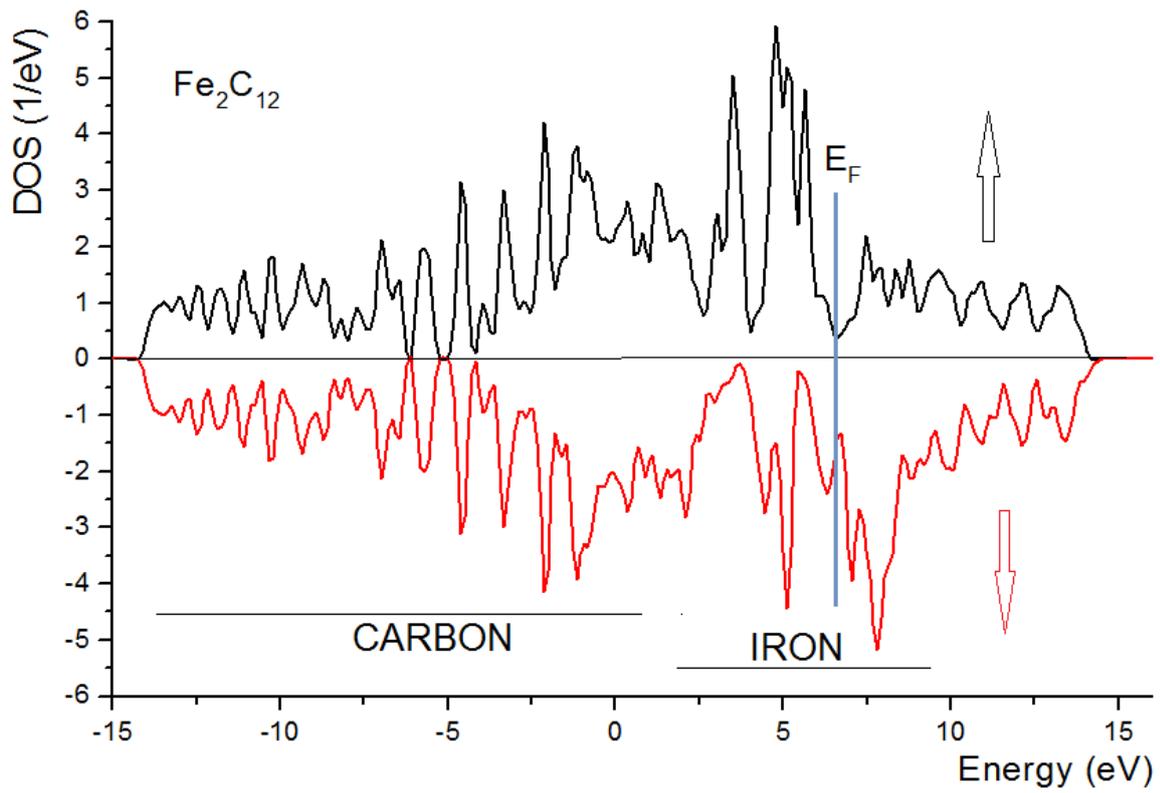

Fig. 5: Fe$_2$C$_{12}$. Spin (↑,↓) projected total density of states DOS.